\documentclass{article}

\usepackage[margin=1in]{geometry}
\usepackage[utf8]{inputenc}
\usepackage[multiple]{footmisc}
\usepackage{amsmath}
\usepackage{amssymb}
\usepackage{amsthm}
\usepackage{bbm}
\usepackage{breakcites}
\usepackage[T1]{fontenc}

\usepackage{multicol}
\usepackage{booktabs}% http://ctan.org/pkg/booktabs
\usepackage{graphicx}
\usepackage{float}
\usepackage{bm}
\usepackage{multirow}
\usepackage{xargs}
\usepackage{listings}
\usepackage{xcolor}
\usepackage{tabularx}
\usepackage{algorithm}
\usepackage{algcompatible}

\usepackage{bbm}
\usepackage{comment}
\usepackage{natbib}
\usepackage[colorlinks=true, allcolors=blue]{hyperref}

\definecolor{codegreen}{rgb}{0,0.6,0}
\definecolor{codegray}{rgb}{0.5,0.5,0.5}
\definecolor{codepurple}{rgb}{0.58,0,0.82}
\definecolor{backcolour}{rgb}{0.95,0.95,0.92}
\lstdefinestyle{mystyle}{
    backgroundcolor=\color{backcolour},   
    commentstyle=\color{codegreen},
    keywordstyle=\color{magenta},
    numberstyle=\tiny\color{codegray},
    stringstyle=\color{codepurple},
    basicstyle=\ttfamily\footnotesize,
    breakatwhitespace=false,         
    breaklines=true,                 
    captionpos=b,                    
    keepspaces=true,                 
    showspaces=false,                
    showstringspaces=false,
    showtabs=false,                  
    tabsize=2
}
\restylefloat{table}

\newcommand{\E}{\mathbb E}

\newcommand{\Var}{\text{Var}}

\newcommand{\forallsp}{\hspace{0.1cm} \forall \hspace{0.1cm}}
\newcommand{\PROCOVATM}{PROCOVA\textsuperscript{TM} }
\newcommand{\PROCOVACMHTM}{PROCOVA\textsuperscript{TM}-CMH }
\theoremstyle{remark}

\makeatletter
\renewcommand*{\ALG@name}{Procedure}
\makeatother
\title{{Prognostic Covariate Adjustment for Binary Outcomes Using Stratification}}
\author{Alyssa M. Vanderbeek$^1$, Jessica L. Ross$^1$, David P. Miller$^1$, Alejandro Schuler$^2$}
\date{
    $^1${Unlearn.AI, Inc., San Francisco, CA}\\%
    $^2${Division of Biostatistics, University of California, Berkeley, Berkeley, CA} \\[2ex]%
    \today
    }

\begin{document}

\maketitle
\begin{abstract}
    
Covariate adjustment and methods of incorporating historical data in randomized clinical trials (RCTs) each provide opportunities to increase trial power. We unite these approaches for the analysis of RCTs with binary outcomes based on the Cochran-Mantel-Haenszel (CMH) test for marginal risk ratio (RR). In \PROCOVACMHTM, subjects are stratified on a single prognostic covariate reflective of their predicted outcome on the control treatment (e.g. placebo). This prognostic score is generated based on baseline covariates through a model trained on historical data. We propose two closed-form prospective estimators for the asymptotic sampling variance of the log RR that rely only on values obtainable from observed historical outcomes and the prognostic model. Importantly, these estimators can be used to inform sample size during trial planning. PROCOVA-CMH demonstrates type I error control and appropriate asymptotic coverage for valid inference. Like other covariate adjustment methods, PROCOVA-CMH can reduce the variance of the treatment effect estimate when compared to an unadjusted (unstratified) CMH analysis. In addition to statistical methods, simulations and a case study in Alzheimer's Disease are given to demonstrate performance. Results show that PROCOVA-CMH can provide a gain in power, which can be used to conduct smaller trials.

\begin{flushleft}
\end{flushleft}
\end{abstract} \hspace{10pt}

\section{Introduction} 
Drug development is a notoriously inefficient, expensive, and time-consuming process \citep{DiMasi,Wong}. Reducing trial sample sizes can speed up trial costs and timelines, but smaller trials can lead to less precision in the treatment effect estimate compared to larger trials, resulting in lower power to identity a truly effective therapy. Larger trials, while able to increase this precision, have trade-offs of additional time required for enrollment and analysis that may delay access to patients in the clinic. Especially for prevalent diseases, this can translate to  millions of additional dollars in spending and the potential for even more patients to receive sub-optimal care. 

In the last decade, there has been increasing discussion about two methodological opportunities to reduce trial sample sizes without loss in power: the leveraging of historical data and covariate adjustment. Historical data is rich with information, but existing methods to incorporate these data into experiments, such as historical borrowing \citep{Ghadessi} or propensity score matching \citep{king_nielsen_2019}, struggle to overcome issues of, for example, bias in results of a trial analysis \citep{thorlund}. There is also little guidance on how to prospectively estimate the magnitude of potential efficiency gains, such as reduced sample size, through incorporation of such data. 

Covariate adjustment offers an opportunity to reduce the variance of the marginal treatment effect, which in turn can reduce sample size without loss of power \citep{FDA_Covar, emaCovar, Hernandez}. Under this framework, there are methods to prospectively estimate potential variance or sample size reduction compared to an analysis that does not use covariate adjustment \citep{Hernandez}. 

To take this one step further, we can unite the incorporation of historical data and covariate adjustment to maximize efficiency gain. For a randomized clinical trial with continuous outcomes, \PROCOVATM leverages historical trial data to produce unbiased marginal treatment effect estimates through covariate adjustment analysis. The information obtained from the historical data alone can be used to predict, in closed-form, potential variance and sample size reductions compared to an unadjusted analysis \citep{Schuler, EMA_2022}. This is achieved through a prognostic model, trained and validated on historical data, that produces a  prognostic score for a subject. This prognostic score is then used as a covariate in PROCOVA's generalized linear model, and by virtue of being a predictor of the outcome, provides an opportunity for variance reduction. The more calibrated these predictions are to the observed outcomes, the greater the potential variance reduction. Expected variance reduction is simply the squared Pearson correlation $R^2$ between the prognostic covariate and the outcome, thus remaining unbiased in the marginal treatment effect estimate \citep{EMA_2022, Schuler}.

Beyond continuous measures, covariate adjustment analysis methods with binary outcomes that provide the same benefits as PROCOVA are also in demand. Logistic regression and the Cochran-Mantel-Haenszel (CMH) test are two commonly used analyses for binary outcomes in randomized trials. When used to estimate marginal risk estimands, combining either of these methods with covariate adjustment can reduce the variance of the treatment effect estimate compared to an unadjusted analysis without loss of power \citep{Ge}. Indeed, the amount of variance reduction possible with these methods is proportional to the predictive ability of the covariate \citep{Richardson,Ge}.

In this paper, we introduce the \PROCOVACMHTM for marginal risk estimands (in particular, risk ratio (RR)) and show that stratifying by the prognostic score can reduce variance of the treatment effect estimate and reduce required sample size. In addition to quantifying this reduction, we present two parsimonious prospective closed-form estimators for the asymptotic sampling variance of the treatment effect that account for the prognostic score. These estimators can be used to prospectively select a trial sample size. Statistical methodology is presented and its performance is demonstrated by simulations and a case study in Alzheimer’s Disease.

\section{Methods} 

The CMH test for binary outcomes stratifies trial subjects into groups based on some baseline covariate such as age or sex. A marginal risk estimand is estimated as a weighted average of the strata risk ratios (RRs) \citep{Noma}. Our focus in this paper is the risk ratio (RR), but the same statistical approach can be used for risk difference (RD).

As a prognostic covariate adjustment method, the PROCOVA-CMH adopts the traditional CMH and stratifies some fixed prognostic score $M$ obtained from a model $m$ trained on historical data. When $m$ is well-calibrated such that the predictions $M$ are highly correlated with the outcome $Y$, we can achieve maximal variance reduction when compared to an unadjusted analysis \citep{Schuler}.

We present asymptotic variance estimators that, under reasonable assumptions, rely only on historical data and fixed trial parameters. These estimators can be used to prospectively select trial sample size and/or quantify the variance reduction possible by using PROCOVA-CMH over an unadjusted (unstratified) analysis. %We also provide approximations for potential variance and sample size reductions when using PROCOVA-CMH over an unadjusted analysis.

\subsection{Notation} \label{sec:notation}

In a two-arm randomized trial of size $n$, let $Y_i \in \{0,1\}$ be a binary outcome, $X_i$ a categorical covariate, and $A_i \in \{0,1\}$ the treatment assignment for subject $i$, $i = {1, \dots, n}$. The covariate $X_i \in \{1, \dots, J\}$ denotes the assignment to one of $J$ strata based on some (baseline) measure, which 
for the purposes of the present method, is independent of randomized treatment assignment. These variables are distributed as follows:
\begin{align*}
    A_i &\overset{IID}{\sim} \text{Bern}(\pi_1) \\
    X_i &\overset{IID}{\sim} \text{Cat}( \boldsymbol{\phi}) %\\
\end{align*}
Here, $\pi_a$ is the randomization probability to arm $a$, independent of stratum assignment and $\text{Cat}$ denotes a categorical distribution with population strata propensity vector $\boldsymbol{\phi} = \{\phi_1,...,\phi_J\}$  such that $\phi_j > 0 \forallsp j$ and $\sum \phi_j = 1$. 

We denote a potential outcome $Y(a)$ on arm $a$. Then ${\mu}_a = E[Y(a)]$ and $\mu_{aj} = E[Y(a)|X=j]$ are the population marginal and stratum-level event probabilities for arm $a$, respectively, such that ${\mu}_a = \sum \mu_{aj} \phi_j$ is a weighted average of the stratum-level probabilities. In the trial, the observed outcome for a subject is $Y_i = Y_i(A)$. Because of randomization, we can identify the causal parameter $\mu_a=E[Y(a)] = E[Y|A=a]$. Then causal inference is possible without further assumptions. The same identification result holds conditionally in each stratum.

Finally, define trial dataset $\boldsymbol{D} = \left\{({X_1}, {A_1}, {Y_1}),...,({X_n}, {A_n}, {Y_n})\right\}$ and the random quantities
\begin{enumerate}
    \item [] $N_{a} = \sum_1^{n} \mathbbm{1}(A_i = a)$ the sample number of subjects on arm $a$;
    \item [] $N_{j} = \sum_1^{n} \mathbbm{1}(X_i = j)$ the sample number of subjects in stratum $j$;
    \item [] $Z_{j} = \sum_1^{n} \mathbbm{1}(X_i = j)$ the sample number of outcomes in stratum $j$;
    \item [] $N_{aj} = \sum_1^{n} \mathbbm{1}(X_i = j)\mathbbm{1}(A_i = a)$ the sample number of subjects on treatment $a$ in stratum $j$; 
    \item [] $Z_{aj} = \sum_1^{n} Y_i \mathbbm 1(X_i = j)\mathbbm{1}(A_i = a)$ the observed number outcomes on arm $a$ in stratum $j$;
\end{enumerate}
By the central limit theorem $N_j$, $N_{aj}$ and $Z_{aj}$ are asymptotically normal with expectations $\phi_j$, $\phi_j \pi_a$, and $\phi_j \pi_a \mu_{aj}$, respectively. Strata outcomes are summarized in a series of $J$ 2x2 tables (Table \ref{tab:sample2x2}). All preceding and following parameter definitions are provided in Appendix Table \ref{appendix:notation-table}.

\begin{table}[H]
\centering
\renewcommand{\arraystretch}{2}
\begin{tabular}{|c|cc|c|}
\hline
   & ${Y=0}$ & ${Y=1}$ & \\ \hline
${A=1}$   & ${n}\phi_j \pi_1 (1-\mu_{1j})$ & ${n}\phi_j \pi_1 \mu_{1j}$ & ${n}\phi_j \pi_1$ \\
${A=0}$   & ${n}\phi_j \pi_0 (1-\mu_{0j})$ & ${n}\phi_j \pi_0 \mu_{0j}$  & ${n}\phi_j \pi_0$ \\ \hline
& & & ${n}\phi_j$ \\
\hline
\end{tabular}
\caption{Expected values 2x2 contingency table on stratum $j$.}
\label{tab:sample2x2}
\end{table}

\subsection{Prognostic Covariate Adjustment} \label{sec:cov-adj}

Suppose we want to adjust the analysis on a prognostic score, which we define as the expected response to the control treatment, conditional on a set of baseline covariates ${B}$: $m(b) = \E[Y(0) \vert B=b]$. Ideally, all information provided by the set of covariates $\boldsymbol{B}$ that is relevant to the outcome is captured in the prognostic score $M$. Since the ideal prognostic score $\mathfrak{M} = m(\cdot)$ is maximally predictive of the actual outcome, it can remove the most amount of variability in the treatment effect estimate when adjusted for (in this case, stratified on). 

In order to estimate this conditional mean, we can apply an algorithm (e.g. deep learning) to a historical dataset. This model $m(\cdot)$ regresses the baseline covariates $B$ on the outcome $Y$ in the historical data. Once $m(\cdot)$ is established, we can apply it in a trial to obtain subject-level prognostic scores based on their baseline covariates, $M_i = m(B_i)$ for subject $i$. And because $m(\cdot)$ is developed through an arbitrary algorithm, it can account for any linear or nonlinear relationships between baseline covariates and the outcome.  

In our setting and for a binary outcome, let $M$ be the estimated conditional probability of observing the outcome estimated by $m(b)$. We will apply $M$ to the PROCOVA-CMH test by assigning patients to a stratum $X$ based on their score through a step function $X = f(M)$:
\begin{align*}
\label{eq:m-bounds}
    X = f(M) = \begin{cases} 1 & \mbox{if } v_{\min} \leq M < v_1 \\
    2 & \mbox{if } v_1 \leq M < v_2 \\
    \vdots & \\
    J & \mbox{if } v_{J-1} \leq M \leq v_{\max}
    \end{cases}
\end{align*}
where $M \in [v_{\min},v_{\max}]$ and $\boldsymbol{v} = \{v_{\min}, v_1, \dots ,v_{J-1}, v_{\max}\}$ represents cutpoints that are chosen \textit{a priori} (e.g. by defining quintiles of the prognostic score in a historical dataset).

\subsection{The CMH test and Mantel-Haenszel Estimator} \label{sec:MH}

We are interested in the marginal risk ratio (RR) $\psi = \frac{\mu_1}{\mu_0}$, which we estimate with the Mantel-Haenszel estimator

\begin{equation}
\hat\psi = 
\frac
{\sum_k Z_{1k} N_{0k} / N_k}
{\sum_k Z_{0k} N_{1k} / N_k} \\
\end{equation}

Under the standard assumption of a homogeneous (common) risk ratio across strata (i.e. $\mu_1/\mu_0 = \mu_{1j}/\mu_{0j}$) this estimator is consistent for the marginal risk ratio.\footnote{When this assumption does not hold, the estimator is consistent for a particular weighted average of the strata RRs (\cite{Noma}).} 

Due to the asymmetric distribution of the risk ratio, it is standard to estimate the variance of the log risk ratio $\log(\psi)$. \cite{Noma} and \cite{Greenland} establish that the asymptotic sample variance of the log Mantel-Haenszel estimator (Appendix \ref{appendix:noma-derivation}) is

\begin{equation}
\label{eq:asymptotic-var}
\Var[\sqrt{n} \log(\hat\psi)] = \sigma_\infty^2 = \frac{\sum_j \phi_j\psi_j[\pi_0{\mu}_{0j} + \pi_1\mu_{1j} - \mu_{0j}\mu_{1j}]}{\psi^2\pi_1\pi_0\left(\sum_j \phi_j\mu_{0j}\right)^2}
\end{equation}

The sample estimator (i.e. using $Z_{aj}/N_{aj}$ for $\mu_{aj}$ etc.) corresponding to the estimand in Eq. \ref{eq:asymptotic-var} is consistent, but can perform poorly in finite samples when some strata are sparsely populated. Thus \cite{Greenland} give the following estimator of the observed sampling variance:% (of $\log(\hat\psi)$):

\begin{equation}
\label{var-estimator}
\hat{\sigma }^2_{\text{GR}} = 
\frac{
\sum _ j 
(N_{1j} N_{0j} Z_j - Z_{1j} Z_{0j} N_j) / N_j^2 }
{ 
\left( \sum_j
Z_{1j} N_{0j} / N_j
\right)
\left( \sum_j
Z_{0j} N_{1j} / N_j
\right) 
}
\end{equation}

This estimator is consistent whether or not the number of strata increase with sample size and whether or not the risk ratio truly is constant across strata \citep{Noma}. 

Together, the point estimate $\hat\psi$ and estimated standard error $\hat\sigma$ can be used to construct Wald-type confidence intervals and p-values ($Z$-test). Since both estimators are $\sqrt{n}$-consistent, strict type-I error control is guaranteed for hypothesis tests in sufficiently large trials.

\subsection{Leveraging historical data to estimate asymptotic variance} \label{sec:var-estimators}

The standard error given by Eq. \ref{var-estimator} is what we use to compute the standard error from trial data. However, to \textit{plan} a trial, we need to estimate the variance \textit{before} we obtain any trial data in order to estimate power and set sample size. Our task is therefore to come up with an alternative variance estimator that uses only historical control data (and assumptions) to estimate the asymptotic sampling variance in the trial. For the purposes of power calculation, the RR $\psi$ is a design parameter - we therefore set it to whatever minimum effect size we wish to target. The randomization fractions $\pi_a$ are similarly set by design.

We present two parsimonious prospective estimators for the asymptotic sampling variance. First, we give a ``plug-in" estimator where all values observed in the historical data are used as point estimates for the paramaters in Eq. \ref{eq:asymptotic-var}. Second, we give a ``modeled" estimator, in which observed values in the historical data and a measure of the calibration of the prognostic model are obtained and some reasonable assumptions about the relationship between strata and outcome probabilities are applied and used in Eq. \ref{eq:asymptotic-var}.

\subsubsection{``Plug-in" estimator}

The plug-in estimator requires specification of all stratum-level quantities, which we may derive from the historical sample under the assumption that the observed values hold for the trial sample. Then $\hat\phi_j = \hat\phi_j^* = \frac{1}{n^*}\sum_i 1(X_i^*=j)$ and $\hat\mu_{0j} = \hat\mu_{0j}^* = \sum_i Y^*_i 1(X^*_i=j)/\sum_i1(X^*_i=j)$. Lastly, we assume a common risk ratio across strata such that $\hat\mu_{1j} = \psi\hat\mu_{0j}$. Plugging these into Eq. \ref{eq:asymptotic-var} and simplifying algebraically gives an estimator:

\begin{equation}
\label{eq:hist-plugin-var}
    {\sigma}^2_{\infty,\text{plug-in}} = %\left(\frac{1}{n-1}\right) 
    \frac{\sum_j \hat{\phi}_j[\pi_0\hat{\mu}_{0j} + \pi_1\hat{\mu}_{1j} - \hat{\mu}_{0j}\hat{\mu}_{1j}]}{{\psi}\pi_0\pi_1\left(\sum_j \hat{\phi}_j \hat{\mu}_{0j}\right)^2}
\end{equation}

%Thus ${\sigma}^2 = f(\boldsymbol{\mu_0}, \boldsymbol{\phi} | \psi,  J, \pi_1)$, where $\boldsymbol{\mu_0}$ and $\boldsymbol{\phi}$ are $J$-length vectors such that $2J + 3$ parameters require specification. 

\subsubsection{``Modeled" estimator}

As an alternative approach to estimating stratum-level values, assume that $\hat{\mu}_{0j}$ and $X$ are linearly related. This allows us to (1) potentially reduce the number of parameters needed from the historical data, and (2) relate the variance of the treatment effect to the coefficient of correlation $r_{XY}$ between $\hat{\mu}_{0j}$ and $X$. %One consequence of (2) is the translation to quantifying efficiency gains over unadjusted analysis that is standard for covariate adjustment analysis. 

Rather than using $\hat{\mu}_{aj}$, we will define modeled probabilities $\Tilde{\mu}_{aj}$ through $\E[{\mu_{aj}}] = {\Tilde{\mu}_{aj}} = g(X) = \beta^{(a)}_0 + \beta^{(a)}_1 X$, $X = \{1,...,J\}$, where 
\begin{align*}
    \beta^{(a)}_1 = r^{(a)}_{XY} \frac{\sigma^{(a)}_{Y}}{\sigma_{X}}, \hspace{1cm} \beta^{(a)}_0 = \hat{\mu}_a - \beta^{(a)}_1 \bar{x}, \hspace{1cm} \sigma^{(a)}_{Y} = \sqrt{{\hat{\mu}}_a(1 - {\hat{\mu}}_a)}, \hspace{1cm} \sigma_{X} = \sqrt{\sum \hat{\phi}_j (j - \bar{x})^2}
\end{align*}
for $r^{(a)}_{XY}$, the Spearman correlation between $X$ and $Y$, and $\bar{x} = \frac{1}{J} \sum j$. Then, $\Tilde{\mu}_a = \frac{1}{J}\sum u_{aj}$ is our modeled (expected) marginal outcome probability on arm $a$. Plugging modeled values into Eq. \ref{eq:asymptotic-var} and simplifying algebraically gives an estimator:
\begin{equation}
\label{eq:var-modeled}
    {\sigma}^2_{\infty,\text{modeled}}   = %\left(\frac{1}{n-1}\right) 
    \frac{\sum_j \Tilde{\phi}_j[\pi_0\Tilde{\mu}_{0j} + \pi_1\Tilde{\mu}_{1j} - \Tilde{\mu}_{0j}\Tilde{\mu}_{1j}]}{{\psi}\pi_0\pi_1\left(\sum_j \Tilde{\phi}_j \Tilde{\mu}_{0j}\right)^2}
\end{equation}
where $\Tilde{\mu}_{1j} = g(X | \hat{\mu}_1) = \beta^{(1)}_0 + \beta^{(1)}_1 X$, $\beta^{(1)}_1 = r^{(1)}_{XY} \frac{\sigma'_{Y}}{\sigma_{X}}$, for $\hat{\mu}_1 = {\psi}\hat\mu_0$, and $\sigma'_{Y} = \sqrt{\hat{\mu}_1(1 - \hat{\mu}_1)}$. While $r^{(0)}_{XY} = r^{*(0)}_{XY}$ can be estimated from the historical data, some assumption must be made about the value of $r^{(1)}_{XY}$. For example, $r^{(0)}_{XY}=r^{(1)}_{XY}$ indicates that the correlation is the same for the control and treatment arms.

For either estimator, the sample variance for a trial with a given sample size $n$ can be prospectively estimated by dividing by $n-1$.

\subsection{Prospective trial planning}
Planning a trial using the PROCOVA-CMH approach consists of three sequential steps: (1) fitting a prognostic model generating prognostic scores for a historical dataset, choosing a number of strata, defining their bounds, and obtaining sample parameters (Procedure \ref{proc:define-strata}); (2) using results from (1) to prospectively estimate the asymptotic sampling variance of the treatment effect (Procedure \ref{proc:prospective-var}); and (3) determining trial sample size (Procedure \ref{proc:samplesize-calc} for the modeled estimator and Procedure \ref{proc:prospective-var-plugin} in the Appendix for the plug-in estimator).

\begin{algorithm}[H]
    \caption{Defining strata and population sample parameters from historical data.}
  \begin{algorithmic}[1]
    \State Fit a prognostic model to historical data to predict control treatment outcomes $M$ for subjects.
    \State Obtain a historical dataset which contains the prognostic scores $M^*$ as generated from the prognostic model.
    \State Subset the historical population sample to align with the inclusion/exclusion criteria of the target trial and define the filtered set as $\boldsymbol{D^*}$. %This will be somewhat limited by the variables contained in $\boldsymbol{D^*}$.
    \State Choose the number of strata $J$. %Multiple values may be evaluated.
    \State Sort the vector $\boldsymbol{M^*}$ and partition the distribution into $J$ non-overlapping intervals (e.g. quintiles, deciles).
    \State Obtain the strata limits $[\text{min}(v_{j}), \text{max}(v_{j})]$ and observed mean prognostic scores.
    \State Estimate sample parameters:
    \begin{itemize}
        \item For the plug-in estimator, calculate stratum-level values $\boldsymbol{\hat{\mu}} = \{\hat{\mu}_{01},...,\hat{\mu}_{0J} \}$ and $\boldsymbol{\hat{\phi}^{}} = \{\hat{\phi}_1,...,\hat{\phi}_J \}$.
        \item For the modeled estimator, calculate Spearman correlation $r^{(0)}_{XY}$, marginal probability $\hat{\mu}_0$, and stratum-level propensities $\boldsymbol{\hat{\phi}^{}} = \{\hat{\phi}_1,...,\hat{\phi}_J \}$.
    \end{itemize} 
  \end{algorithmic}  
  \label{proc:define-strata}
\end{algorithm}

\begin{algorithm}[H]
    \caption{Using the modeled estimator for prospectively estimate variance.}
  \begin{algorithmic}[1]
    \State Use Procedure \ref{proc:define-strata} to obtain estimates  $\hat{\mu}^{}_0$, $\boldsymbol{\hat{\phi}^{}} = \{\hat{\phi}_1,...,\hat{\phi}_J \}$, and $r^{(0)}_{XY}$ from historical data $\boldsymbol{D^*}$.
    \State Specify trial parameters ${\psi}$ and $\pi_1$.
    \State Calculate estimates of $\boldsymbol{\Tilde{\mu}_0} = \{\Tilde{\mu}_{01},...,\Tilde{\mu}_{0J}\}$.
    \begin{itemize}
        \item Calculate $\beta^{(0)}_1 = r^{(0)}_{XY} \frac{\sqrt{\hat{\mu}^{}_0(1 - \hat{\mu}^{}_0)}}{\sqrt{\sum \hat{\phi}^{}_j (j - \bar{x})^2}}$, where $\bar{x} = \frac{1}{J}\sum x$.
        \item Set $\beta^{(0)}_0 = \hat{\mu}_0 - \beta^{(0)}_1 \bar{x}$.
        %\item Calculate $\beta_0 = \bar{\mu}_0 - \beta_1 \bar{j}$. This is a simple trick to fix the marginal probability close to $\bar{p}_0$.
        \item Calculate $\Tilde{\mu}_{0j} = \beta^{(0)}_0 + \beta^{(0)}_1 x$.% for all strata ($j = 1,...,5$). 
        \item Define ${\Tilde{\mu}}_0 = \frac{1}{J}\sum \Tilde{\mu}_{0j}$
    \end{itemize}
    \State Repeat Step 3 for $\boldsymbol{\Tilde{\mu}_1}$, where $\hat{\mu}_1 = {\psi} \hat{\mu}^{}_0$.
    \State Plug all values into Eq. \ref{eq:var-modeled} to estimate the asymptotic variance.
  \end{algorithmic}
\label{proc:prospective-var}
\end{algorithm}

\begin{algorithm}[H]
    \caption{Determining sample size for a prospective trial.}
  \begin{algorithmic}[1]
    \State Use Procedure \ref{proc:define-strata} to obtain relevant parameter estimates from historical data $\boldsymbol{D^*}$.
    \begin{itemize}
        \item  For the plug-in estimator, obtain stratum-level values $\boldsymbol{\hat{\mu}} = \{\hat{\mu}_{01},...,\hat{\mu}_{0J} \}$ and $\boldsymbol{\hat{\phi}^{}} = \{\hat{\phi}_1,...,\hat{\phi}_J \}$.
        \item For the modeled estimator, obtain Spearman correlation $r^{(0)}_{XY}$, marginal probability $\hat{\mu}_0$, and stratum-level propensities $\boldsymbol{\hat{\phi}^{}} = \{\hat{\phi}_1,...,\hat{\phi}_J \}$
    \end{itemize}
        \State For the prospective trial, specify the desired power, significance level $\alpha$, treatment effect (risk ratio) $\psi$, randomization probability $\pi$, and other parameters relevant to the sample size calculation (e.g. anticipated dropout rate). 
    \State Estimate the asymptotic sampling variance $\sigma^2_\infty$ for the log risk ratio, $\log(\psi)$, using the plug-in (Eq. \ref{eq:hist-plugin-var}) and/or modeled estimator (Eq. \ref{eq:var-modeled}, Procedure \ref{proc:prospective-var}).
    \State Calculate minimum required sample size to achieve the desired power where the probability of a two-sided $p$-value being less than $\alpha$ is asymptotically
    \begin{equation}
        \Phi \left( \Phi^{-1}(\alpha/2) + \sqrt{n}\frac{\tau}{\sigma} \right) + \Phi \left( \Phi^{-1}(\alpha/2) - \sqrt{n}\frac{\tau}{\sigma} \right)
    \end{equation}
    where $\tau = \log(\psi)$ (chosen in Step 2) and $\sigma$ is the square root of the asymptotic sampling variance estimated in Step 3. 
  \end{algorithmic}
\label{proc:samplesize-calc}
\end{algorithm}

In addition to prospectively estimating variance and calculating sample size directly, we can quantify the gains from PROCOVA-CMH when compared to an unadjusted CMH test. As a covariate adjustment method, PROCOVA-CMH can reduce the variance of the treatment effect estimate compared to an unadjusted analysis (in this setting, an unstratified CMH test) \citep{Hernandez,Schuler}. 

Variance reduction is simply defined as $\gamma = 1 - \frac{\sigma^2_{\text{hist,PROCOVA-CMH}}}{\sigma^2_{\text{hist,unadj}}}$, where estimates of $\sigma^2$ can be obtained from the plug-in or modeled estimators (Appendix \ref{appendix:var_ratio}). This value translates to increased power which can be used to reduce required sample size.

\section{Simulations} \label{sec:sims}

Our simulations evaluate the performance of the PROCOVA-CMH compared to an unadjusted analysis and the accuracy of the plug-in and modeled estimators. For our data generation procedure, we assume that the distribution of prognostic scores (expected response probabilities to the control) follows a truncated (0,1) Normal distribution. This corresponds to a common scenario where, for example, most people are of average risk of the outcome and a smaller proportion are of high and low risk. To define population outcome probabilities in each strata, we calculate the mean value within each of $J$ strata. In our simulation scenario, we define quintiles ($J=5$). The mean of the truncated Normal distribution is calculated to align with a pre-specified marginal response probability $\mu_0$ and a standard deviation value is chosen to approximate from correlation $r_{XY}$. Outcome probabilities for the treatment arm are defined in the same way where $\mu_1 = \psi\mu_0$.

For one simulation, we generate a historical dataset $\boldsymbol{D^{*}}$ of 10,000 control samples $(X^{*},Y^{*})$ using the values defined for $\mu_0$. The same data generation procedure is used for a trial dataset $\boldsymbol{D}$ of 800 samples. Each subject is assigned to a stratum with probability $\phi_j$, and their potential outcomes $Y_i(0)$ and $Y_i(1)$ are drawn from $Y_i(0)|X \sim \text{Bern}(\mu_{0j})$ and $Y_i(1)|X \sim \text{Bern}(\mu_{1j})$. Subjects are then assigned to the treatment arm with fixed probability $\pi_1$, and we take their observed outcome as $Y_i = Y_i(A_i)$. Finally, we run a trial and perform both a PROCOVA-CMH and unadjusted test. Simulations are performed 1,000 times in R v.4.1.2 (R Core Team, 2021). 

We assess performance of three components: 
\begin{enumerate}
    \item Properties of the estimates, including mean-squared-error (MSE) of the marginal RR estimate, the asymptotic coverage of the marginal RR estimate, and the rejection rate based on a two-sided t-test.
    \item Prospective sample variance is estimated from as $\frac{1}{n-1}\sigma^2$ using both the plug-in (Eq. \ref{eq:hist-plugin-var}) and modeled (Eq. \ref{eq:var-modeled}) estimators, and compared to the observed sampling variance. Performance is measured by the average bias of the estimate.
    \item The predicted variance reduction of using PROCOVA-CMH vs. an unadjusted test, as estimated from our estimators, compared to the observed variance reduction. We present the mean observed value and average estimator bias, as well as the mean values of $r^2_{XY}$ from the historical and trial datasets.
\end{enumerate}
We consider 6 scenarios under different population values and distributional shifts between historical and trial populations. In the baseline scenario, we consider $\mu_0=0.5$ and $r^2_{XY} \approx 0.2$. In continuous PROCOVA, for example, $r^2_{XY} \approx 0.2$ is associated with a 20\% variance reduction or 20\% decrease in the required overall trial sample size. In scenario 2, we consider a shift in $r^2_{XY}$ between historical and trial data to represent differences in model calibration. In scenario 3, we specify a shift in $\mu_0$ between historical and trial data to represent a case where the historical data underestimates the observed outcome in the trial sample. We also consider a scenario with smaller $r^2_{XY}$ (scenario 4) and a scenario with both smaller $\mu_0$ and $r^2_{XY}$ (scenario 5). Finally, we look at a scenario with 2:1 randomization (scenario 6). Values in red in Table \ref{tab:sim-scenarios-main} denote deviations from the baseline scenario. Correlations in the trial control and treatment arms are assumed equal. In all scenarios we test under the null hypothesis ${\psi}=1$ and alternatives where ${\psi}=0.75$ and ${\psi}=1.25$ (Table \ref{tab:sim-scenarios-main}). %Simulations scenarios in the case of a non-Gaussian dependence structure between $X$ and $\mu$ are provided in Appendix \ref{appendix:copula-sims}.

\begin{table}[H]
    \centering
    %\resizebox{\columnwidth}{!}{
    \begin{tabular} {|l|c|c|c|c|c|c|}
    \hline
         Scenario & J & ${\mu}_0$ & ${\mu}^{(\boldsymbol{D})}_0 - {\mu}^{(\boldsymbol{D*})}_0$ & $r^2_{XY}$ & $r^{2(\boldsymbol{D})}_{XY} - r^{2(\boldsymbol{D*})}_{XY}$ & $\pi_1:\pi_0$ \\
          \hline
         \textbf{1. Baseline} & 5 & 0.5 & 0 & 0.2 & 0 & 1:1 \\  
         \textbf{2. Shift in $r^2_{XY}$} & 5 & 0.5 & 0 & 0.2 & \textcolor{red}{-0.1} & 1:1 \\  
         \textbf{3. Shift in ${\mu}_0$} & 5 & \textcolor{red}{0.3} & \textcolor{red}{0.1} & 0.2 & 0 & 1:1 \\  
         \textbf{4. Smaller $r^2_{XY}$} & 5 & 0.5 & 0 & \textcolor{red}{0.1} & 0 & 1:1\\  
         \textbf{5. Smaller ${\mu}_0$ and $r^2_{XY}$} & 5 & \textcolor{red}{0.3} & 0 & \textcolor{red}{0.1} & 0 & 1:1 \\  
         \textbf{6. 2:1 randomization} & 5 & 0.5 & 0 & 0.2 & 0 & \textcolor{red}{2:1} \\  
         \hline
    \end{tabular}
    %}
    \caption{Simulation scenarios. Squared correlation values are approximate, with corresponding truncated normal standard deviations 0.29 (for $r^2_{XY}=0.2$) and 0.16 (for $r^2_{XY}=0.1$).}
    \label{tab:sim-scenarios-main}
\end{table}

Stratifying on the prognostic score produces an estimate of the log RR with similar asymptotic coverage (unbiasedness), type I error, smaller MSE, and greater power compared to the unadjusted analysis (Table \ref{tab:sims_asymptotics-main}). The increase in power is due to a reduction in the variance of the estimated treatment effect. Notably, the amount of variance reduction is approximately equal to the $r^2_{XY}$ observed in the control arm of the trial (and assuming, as we have, that the correlation is the same in the treatment arm). It is also worth noting that the amount of variance reduction is also a function of other parameters, namely the marginal outcome probability, the treatment effect, and the randomization ratio (Table \ref{tab:bias-var-reduction-main}). This result reflects our variance estimators (and their subsequent ratio in estimating variance reduction) and the results of other work exploring gains of covariate adjustment \citep{Borm,Siegfried,Schuler}.

\begin{table}[H]
    \centering
    \resizebox{\columnwidth}{!}{
\begin{tabular}{|l|cc|cc|cc|}
    \hline
    \multicolumn{1}{|c|}{ } & \multicolumn{2}{c|}{\textbf{MSE}} & \multicolumn{2}{c|}{\textbf{Rejection rate}} & \multicolumn{2}{c|}{\textbf{Coverage}} \\
    \cline{2-3} \cline{4-5} \cline{6-7}
     & PROCOVA-CMH & Unadjusted & PROCOVA-CMH & Unadjusted & PROCOVA-CMH & Unadjusted\\
    \hline
    \multicolumn{7}{l}{\textbf{Scenario 1}}\\
    \hline
    ${\psi}=1$ & 0.0041 & 0.0052 & 0.057 & 0.055 & 0.945 & 0.947\\
    \hline
    ${\psi}=0.75$ & 0.0046 & 0.0054 & 0.899 & 0.821 & 0.900 & 0.913\\
    \hline
    ${\psi}=1.25$ & 0.0059 & 0.0072 & 0.949 & 0.900 & 0.930 & 0.934\\
    \hline
    \multicolumn{7}{l}{\textbf{Scenario 2}}\\
    \hline
    ${\psi}=1$ & 0.0050 & 0.0054 & 0.058 & 0.063 & 0.943 & 0.936\\
    \hline
    ${\psi}=0.75$ & 0.0035 & 0.0039 & 0.942 & 0.912 & 0.947 & 0.947\\
    \hline
    ${\psi}=1.25$ & 0.0059 & 0.0066 & 0.973 & 0.961 & 0.952 & 0.922\\
    \hline
    \multicolumn{7}{l}{\textbf{Scenario 3}}\\
    \hline
    ${\psi}=1$ & 0.0059 & 0.0073 & 0.051 & 0.050 & 0.948 & 0.949\\
    \hline
    ${\psi}=0.75$ & 0.0081 & 0.0094 & 0.669 & 0.548 & 0.854 & 0.860\\
    \hline
    ${\psi}=1.25$ & 0.0089 & 0.0106 & 0.816 & 0.748 & 0.915 & 0.922\\
    \hline
    \multicolumn{7}{l}{\textbf{Scenario 4}}\\
    \hline
    ${\psi}=1$ & 0.0050 & 0.0054 & 0.058 & 0.063 & 0.943 & 0.936\\
    \hline
    ${\psi}=0.75$ & 0.0035 & 0.0039 & 0.942 & 0.912 & 0.947 & 0.947\\
    \hline
    ${\psi}=1.25$ & 0.0059 & 0.0066 & 0.973 & 0.961 & 0.952 & 0.922\\
    \hline
    \multicolumn{7}{l}{\textbf{Scenario 5}}\\
    \hline
    ${\psi}=1$ & 0.0106 & 0.0116 & 0.042 & 0.051 & 0.959 & 0.953\\
    \hline
    ${\psi}=0.75$ & 0.0079 & 0.0086 & 0.538 & 0.477 & 0.946 & 0.952\\
    \hline
    ${\psi}=1.25$ & 0.0141 & 0.0156 & 0.707 & 0.679 & 0.948 & 0.950\\
    \hline
    \multicolumn{7}{l}{\textbf{Scenario 6}}\\
    \hline
    ${\psi}=1$ & 0.0046 & 0.0060 & 0.056 & 0.054 & 0.942 & 0.945\\
    \hline
    ${\psi}=0.75$ & 0.0045 & 0.0053 & 0.890 & 0.801 & 0.920 & 0.921\\
    \hline
    ${\psi}=1.25$ & 0.0066 & 0.0084 & 0.916 & 0.840 & 0.923 & 0.935\\
    \hline
\end{tabular}
}
    \caption{Mean-squared error of the treatment effect estimate, rejection rate, and coverage results for the PROCOVA-CMH approach vs. unadjusted analysis. Type I error is given by the rejection rate under the null hypothesis, and power is given for both alternative hypotheses.}
    \label{tab:sims_asymptotics-main}
\end{table}

The plug-in and modeled variance estimators are approximately unbiased when the historical and trial populations are equal (i.e. the marginal outcome probability is the same and the prognostic model is equally calibrated such that $r^{*2}_{XY}=r^2_{XY}$) (scenarios 1, 4, 5, 6, Table \ref{tab:bias-var-main}). However, some bias exists when there is a population shift (scenarios 2 and 3). In scenario 2, there is a difference in model calibration ($r^2_{XY}$ in the trial is smaller than in the historical data). The estimators are unbiased for the unadjusted test, but slightly underestimate the variance of the treatment effect using PROCOVA-CMH, on average. In scenario 3 where we have a population shift in the marginal outcome probability, there is substantial bias for both estimators for the PROCOVA-CMH and unadjusted tests. In all, the estimators are very similar, though the modeled estimator produces an estimate of the variance with a smaller or equal absolute bias compared to the plug-in estimator (Table \ref{tab:bias-var-main}).

\begin{table}[H]
    \centering
    %\resizebox{\columnwidth}{!}{
\begin{tabular}{|l|cc|cc|}
\hline
\multicolumn{1}{|c|}{ } & \multicolumn{2}{c|}{\textbf{PROCOVA-CMH}} & \multicolumn{2}{c|}{\textbf{Unadjusted analysis}} \\
\cline{2-5} 
 & Modeled & Plug-in & Modeled & Plug-in \\
\hline
\multicolumn{5}{l}{\textbf{Scenario 1}}\\
\hline
${\psi}=1$ & 0.0000 & -0.0001 & 0.0000 & 0.0000 \\
\hline
${\psi}=0.75$ & 0.0002 & 0.0005 & 0.0003 & 0.0003 \\
\hline
${\psi}=1.25$ & -0.0001 & -0.0003 & -0.0001 & -0.0001 \\
\hline
\multicolumn{5}{l}{\textbf{Scenario 2}}\\
\hline
${\psi}=1$ & -0.0006 & -0.0006 & 0.0000 & 0.0000 \\
\hline
${\psi}=0.75$ & -0.0008 & -0.0005 & 0.0000 & 0.0000 \\
\hline
${\psi}=1.25$ & -0.0005 & -0.0006 & 0.0000 & 0.0000 \\
\hline
\multicolumn{5}{l}{\textbf{Scenario 3}}\\
\hline
${\psi}=1$ & 0.0025 & 0.0025 & 0.0029 & 0.0029 \\
\hline
${\psi}=0.75$ & 0.0033 & 0.0037 & 0.0038 & 0.0039 \\
\hline
${\psi}=1.25$ & 0.0021 & 0.0018 & 0.0024 & 0.0024 \\
\hline
\multicolumn{5}{l}{\textbf{Scenario 4}}\\
\hline
${\psi}=1$ & 0.0000 & 0.0000 & 0.0000 & 0.0000 \\
\hline
${\psi}=0.75$ & 0.0000 & 0.0001 & 0.0000 & 0.0000 \\
\hline
${\psi}=1.25$ & 0.0000 & -0.0001 & 0.0000 & 0.0000 \\
\hline
\multicolumn{5}{l}{\textbf{Scenario 5}}\\
\hline
${\psi}=1$ & -0.0001 & -0.0001 & -0.0001 & -0.0001 \\
\hline
${\psi}=0.75$ & 0.0001 & 0.0003 & 0.0002 & 0.0002 \\
\hline
${\psi}=1.25$ & -0.0001 & -0.0003 & -0.0001 & -0.0001 \\
\hline
\multicolumn{5}{l}{\textbf{Scenario 6}}\\
\hline
${\psi}=1$ & -0.0001 & -0.0001 & 0.0000 & 0.0000 \\
\hline
${\psi}=0.75$ & 0.0001 & 0.0004 & 0.0002 & 0.0002 \\
\hline
${\psi}=1.25$ & -0.0001 & -0.0003 & -0.0001 & -0.0001 \\
\hline
\end{tabular}
%}
    \caption{Average bias of the variance estimate for the plug-in and modeled estimators, compared to the observed sampling variance of the trial.}
    \label{tab:bias-var-main}
\end{table}

In addition to the estimates of the variance itself, we also examine bias of the estimated variance reduction (Table \ref{tab:bias-var-reduction-main}). In all but two instances across the 6 scenarios (Scenarios 2 and 3), the estimated variance reduction using the modeled variance estimator has smaller absolute bias compared to the plug-in estimator. Additionally, in scenario 2, the magnitude of bias in the estimated variance reduction is proportional to the population shift in $r^2_{XY}$ (Table \ref{tab:bias-var-reduction-main}).

%It's important to note that variance and variance reduction (and power gain) have a direct impact on the required sample size for a trial. 

\begin{table}[H]
    \centering
    \resizebox{\columnwidth}{!}{
\begin{tabular}{|l|c|cc|cc|}
\hline
\multicolumn{1}{|c|}{ } & \multicolumn{1}{c|}{\textbf{Variance reduction}} & \multicolumn{2}{c|}{\textbf{Bias}} & \multicolumn{2}{c|}{$\boldsymbol{R^2}$} \\
\cline{2-6} 
 & Observed & Modeled estimator & Plug-in estimator & Historical & Trial shift \\
\hline
\multicolumn{6}{l}{\textbf{Scenario 1}}\\
\hline
${\psi}=1$ & 0.2047 & \textbf{0.0027} & 0.0034 & 0.2074 & 0.0007 \\
\hline
${\psi}=0.75$ & 0.1946 & \textbf{0.0062} & -0.0385 & 0.2074 & 0.0007 \\
\hline
${\psi}=1.25$ & 0.1951 & \textbf{0.0057} & 0.0436 & 0.2074 & 0.0007\\
\hline
\multicolumn{6}{l}{\textbf{Scenario 2}}\\
\hline
${\psi}=1$ & 0.0868 & \textbf{0.1206} & 0.1213 & 0.2047 & -0.1162 \\
\hline
${\psi}=0.75$ & 0.0848 & 0.1160 & \textbf{0.0713} & 0.2047 & -0.1162 \\
\hline
${\psi}=1.25$ & 0.0851 & \textbf{0.1156} & 0.1536 & 0.2047 & -0.1162 \\
\hline
\multicolumn{6}{l}{\textbf{Scenario 3}}\\
\hline
${\psi}=1$ & 0.1993 & -0.0166 & \textbf{-0.0129} & 0.1827 & 0.0201 \\
\hline
${\psi}=0.75$ & 0.1879 & \textbf{-0.0127} & -0.0389 & 0.1827 & 0.0201 \\
\hline
${\psi}=1.25$ & 0.1991 & \textbf{-0.0193} & 0.0202 & 0.1827 & 0.0201 \\
\hline
\multicolumn{6}{l}{\textbf{Scenario 4}}\\
\hline
${\psi}=1$ & 0.0868 & \textbf{0.0030} & 0.0038 & 0.0898 & 0.0014 \\
\hline
${\psi}=0.75$ & 0.0848 & \textbf{0.0022} & -0.0168 & 0.0898 & 0.0014 \\
\hline
${\psi}=1.25$ & 0.0851 & \textbf{0.1313} & 0.1484 & 0.0898 & 0.0014 \\
\hline
\multicolumn{6}{l}{\textbf{Scenario 5}}\\
\hline
${\psi}=1$ & 0.0902 & \textbf{0.0028} & 0.0036 & 0.0930 & 0.0035 \\
\hline
${\psi}=0.75$ & 0.0871 & \textbf{0.0042} & -0.0114 & 0.0930 & 0.0035 \\
\hline
${\psi}=1.25$ & 0.0882 & \textbf{0.0035} & 0.0213 & 0.0930 & 0.0035 \\
\hline
\multicolumn{6}{l}{\textbf{Scenario 6}}\\
\hline
${\psi}=1$ & 0.2011 & \textbf{0.0062} & 0.0070 & 0.2074 & -0.0010 \\
\hline
${\psi}=0.75$ & 0.2049 & \textbf{0.0134} & -0.0352 & 0.2074 & -0.0010 \\
\hline
${\psi}=1.25$ & 0.1789 & \textbf{0.0070} & 0.0461 & 0.2074 & -0.0010 \\
\hline
\end{tabular}
}
    \caption{Results for the variance reduction of using PROCOVA-CMH over an unadjusted analysis. Observed variance reduction, bias of the modeled and plug-in variance reduction estimators, and historical and trial $R^2$. Bolded values indicate smaller absolute bias.}
    \label{tab:bias-var-reduction-main}
\end{table}

\section{Case study: Alzheimer's Disease}

We apply the PROCOVA-CMH procedure to a phase III RCT of Docosahexaenoic Acid (DHA) supplementation in Alzheimer’s Disease (AD) (NCT00440050) \citep{Quinn}. The DHA trial enrolled $n = 402$ subjects at a 3:2 randomization scheme, assigning 164 to placebo and
238 to the experimental treatment. We artificially define a binary outcome as disease progression (yes/no) characterized by a 12-month change in the Alzheimer’s Disease Assessment Scale-Cognitive Subscale (ADAS-Cog 11) exceeding some threshold $c$. Two thresholds are considered: $c=4$ and $c=8$, where a cutoff of 4 may be considered clinically relevant for signifying worsening disease \citep{Rockwood}, and a cutoff of $8$ yields a right-skewed distribution of $M$. All analysis was performed in R 4.1.2. (2021) using the 'metafor' package \citep{metafor}.

To generate prognostic scores and define the strata for this case study, we utilized a generative model that we previously developed for Alzheimer’s Disease progression built on historical data from the Alzheimer’s Disease Neuroimaging Initiative (ADNI) and the C-Path Online Data Repository for Alzheimer’s Disease \citep{Bertolini}. A newer version of this model, used here, adds important secondary endpoints and makes dataset and architectural refinements. Both versions used the same integrated dataset from 6,919 subjects, covering 64 background and longitudinal variables. Panel data from both clinical and observational studies in Mild Cognitive Impairment (MCI) and AD patients were used to develop the model, wherein patients with MCI make up about 25\% of the subject population. The integrated dataset was split into training, validation, and test datasets at a 0.5 : 0.2 : 0.3 ratio. Baseline covariates including demographic characteristics (sex, age, education level, etc), lab tests (APOE status, neuroimaging results, etc), vital signs (blood pressure, weight, etc), and secondary outcome assessments (Clinical Dementia Rating, Neuropsychiatric Inventory, etc) were used to generate prognostic scores.

We now apply PROCOVA-CMH. After filtering the test dataset to patients with baseline MMSE between 14 and 26 (inclusive) to reflect the inclusion/exclusion criteria of the DHA trial \citep{Quinn}, 2,012 subjects who received placebo remained. The test dataset is what we call $\boldsymbol{D^*}$. From this data we defined limits, correlation, marginal outcome probability, and propensities for 5 strata (number of strata chosen for illustrative purposes) (Procedure \ref{proc:define-strata}). In application to the DHA trial, and using Procedure \ref{proc:prospective-var}, we estimated the sampling variance for both the PROCOVA-CMH and unadjusted tests from the historical data $\boldsymbol{D^*}$ using the plug-in and modeled prospective variance estimators (Section \ref{sec:var-estimators}) and the trial sample size. Estimates for potential variance reduction were also computed. We conducted both a PROCOVA-CMH and unadjusted test, noting that 92 subjects (32 control, 60 treatment) enrolled in this trial had missing outcome data (treated as NA in the analysis). We report the observed treatment effects estimates, variance estimates, and variance reduction observed by using the PROCOVA-CMH vs. the unadjusted test. We compare our historically-informed prospective estimates of sampling variance and variance reduction to what was observed in analysis. 

\begin{table}[H]
    \centering
    %\resizebox{\columnwidth}{!}{
\begin{tabular}{|c|ccc|cc||} 
\hline
\multicolumn{1}{|c|}{ } & \multicolumn{3}{c|}{$\hat{r}^2_{XY}$} & \multicolumn{2}{c|}{$\hat{\mu}_0$} \\%& \multicolumn{3}{c|}{Variance reduction (\%)} \\
\cline{2-4} \cline{5-6} %\cline{7-9}
  & Historical & DHA
(control)  & DHA
(treatment) & Historical & DHA \\%& 'Plug-in' & 'Modeled' & Observed\\
\hline
$c=4$ & 0.079 & 0.045 & 0.082 & 0.388 & 0.474 \\%& 9.1 & 7.1 & 6.2\\
\hline
$c=8$ & 0.063 & 0.091 & 0.095 & 0.187 & 0.239 \\%& 6.3 & 4.7 & 8.1\\
\hline
\end{tabular}
%}
\caption{Correlation between strata and outcome in historical and trial data and observed marginal outcome probabilities for the control treatment in historical and trial data.}
    \label{tab:dha-data-vals}
\end{table}

\begin{table}[H]
    \centering
    \resizebox{\columnwidth}{!}{
\begin{tabular}{|c|ccc|cc|ccc|} 
\hline
\multicolumn{1}{|c|}{ } & \multicolumn{3}{c|}{Prospective variance estimates} & \multicolumn{2}{c|}{Observed variance estimates} & \multicolumn{3}{c|}{Variance reduction (\%)} \\
\cline{2-4} \cline{5-7} \cline{7-9}
  & Plug-in & Modeled & Unadjusted
& PROCOVA-CMH & Unadjusted & Plug-in & Modeled & Observed\\
\hline
$c=4$ & 0.015 & 0.017 & 0.019 & 0.013 & 0.014 & 9.1 & 7.1 & 6.2\\
\hline
$c=8$ & 0.037 & 0.043 & 0.045 & 0.038 & 0.041 & 6.3 & 4.7 & 8.1\\
\hline
\end{tabular}
}
\caption{Predicted and observed variance estimates and variance reduction. Predicted values are given for both the 'plug-in' and 'modeled' variance estimators.}
    \label{tab:var-vals}
\end{table}

%Variance reduction was predicted from Procedure \ref{proc:prospective-var} based on the values obtained from the historical data and assuming a null treatment effect ${\psi}=1$ \textcolor{red}{(Can use something else.)}.%

\begin{table}[H]
    \centering
    \begin{tabular}{|c|c|c|}
\hline
  & PROCOVA-CMH & Unadjusted\\
\hline
$c=4$ & $0.795 \pm 0.115$ & $0.772 \pm 0.118$\\
\hline
$c=8$ & $0.952 \pm 0.195$ & $0.921 \pm 0.204$\\
\hline
\end{tabular}
    \caption{Estimate of the marginal risk ratio and standard errors.}
    \label{tab:cmh-estimates}
\end{table}

In both cutoff scenarios, the prospective plug-in variance estimator provides a close estimate of the observed variance estimate in the trial using PROCOVA-CMH (0.015 predicted vs. 0.013 observed for $c=4$ and 0.037 vs. 0.038 observed for $c=8$). By contrast, the modeled variance estimator overestimates the observed variance which, if used in trial planning, can result in an underpowered study (sample size is too small to detect a treatment effect at a pre-specified power level). For the unadjusted analysis, both estimators (equivalent) overestimate the observed variance (Table \ref{tab:var-vals}). 

With respect to variance reduction, effect of shifts in $(\hat{r}^{2*}_{XY}, \hat{r}^2_{XY})$ and $(\hat{\mu}^*_0, \hat{\mu}_0)$ are reflected in the accuracy of the variance reduction prediction (Table \ref{tab:var-vals}). For cutoff $c=4$, the average $\hat{r}^2_{XY}$ between treatment arms shifts from the historical $\hat{r}^2_{XY}$ by approximately 0.015 (Table \ref{tab:dha-data-vals}), and the modeled variance reduction estimate based on the historical values overestimates the observed variance reduction by 0.009 (Table \ref{tab:var-vals}). Additionally, the observed variance reduction (6.2\%) is almost exactly the average trial $\hat{r}^2_{XY}$ (0.063) (Tables \ref{tab:dha-data-vals} and \ref{tab:var-vals}). By comparison, the plug-in variance reduction estimate overestimates the observed reduction by 0.029. Meanwhile the cutoff $c=8$, the modeled estimate of variance reduction underestimates the observed by 0.034 compared to 0.018 for the plug-in estimate.

Finally, the PROCOVA-CMH estimate of the marginal risk ratio is consistent for the estimate of the unadjusted test (Table \ref{tab:cmh-estimates}).

\section{Discussion}
We’ve presented a method of stratified covariate adjustment for binary outcomes in randomized trials that leverages historical data and machine learning. Our results show that PROCOVA-CMH provides a consistent treatment effect estimate and can reduce the estimated variance when compared to an unadjusted analysis. Of note, one is able to prospectively estimate the variance based only on information obtained from historical data and trial specifications, which allows for sample size selection. The potential variance reduction obtainable by using PROCOVA-CMH over an unadjusted analysis can be similarly estimated. 

PROCOVA-CMH is model-agnostic, so the choices of the prognostic score and model algorithm are at the will of the user. By being trained on other covariates, and can thus account for any linear or nonlinear relationships between those covariates and the outcome. Ideally, all information provided by the set of covariates that is relevant to the outcome is captured in the prognostic score. As previously shown \citep{Schuler}, variance and sample size reductions are maximized when the generated prognostic score is maximally correlated with the outcome. However, in large-enough samples there is no risk of type I error or loss in power relative to unadjusted analysis if the model used provides inaccurate predictions or prognostic scores with low correlation to the outcome. In terms of metrics for the prognostic score, we use the Spearman correlation $r_{XY}$ to measure association, but other measures such as the area under the receiver operating characteristic curve (AUC) may be alternatively used for a binary outcome to calibrate a model where the goal is classification. Numerical approximations between the AUC and $r_{XY}$ can be used in tandem in this case to develop the model to provide maximal potential gains with PROCOVA-CMH \citep{gneiting}.

We give two estimators for the variance that use information obtained from historical data: what we call the plug-in estimator, which directly inputs the strata-level values observed in an relevant historical dataset into the formula (Eq. \ref{eq:hist-plugin-var}), and a modeled estimator, which models expected strata-level values based on the model calibration (Spearman $r_{XY}$), historical values, and some reasonable assumptions (Eq. \ref{eq:var-modeled}). Each has their own advantages and drawbacks; the plug-in estimator requires fewer calculations for the analyst. However, this approach does not directly relate the calibration of the prognostic model to the expected variance of the estimate or expected degree of variance reduction for using PROCOVA-CMH. Conversely, the modeled estimator requires a few additional calculations from the analyst, but may be less biased in its prediction of the variance reduction and allows a direct connection to prognostic model calibration. 

Although these estimators demonstrate desirable properties, they can be biased if there are shifts between the historical and trial populations. This bias is largest when the shift occurs in the calibration of the prognostic model, but was also shown in the case where the marginal outcome probability deviates (Tables \ref{tab:bias-var-main} and \ref{tab:bias-var-reduction-main}).  These properties were also reflected in our case study. Still, parameters can be varied (e.g. down-weighting the Spearman $r_{XY}$) to prevent overly optimistic estimates if the analyst planning a prospective trial expects any population shifts.

Logistic regression is another commonly embraced analysis method for binary outcomes, and can be desirable in its  ability to account for multiple covariates. It may be possible to achieve a similar level of variance reduction to PROCOVA-CMH from a multiple covariate adjusted logistic regression for marginal estimands. Future work exploring logistic regression adjusted for prognostic scores in a similar fashion described here can build on the existing body of knowledge \citep{Richardson,Ge,Neuhaus}.

In sum, PROCOVA-CMH is a method of prognostic covariate adjustment that directly incorporates historical data and leverages prognostic model development in a way that can be used to power prospective randomized trials.  As the predictive power of the prognostic score is only dependent on the model used to generate it, the continued refinement of machine learning methods and growing availability of high-quality, high-dimensional datasets suggest that more efficient treatment effect estimation for binary outcomes is increasingly possible.

%%%%%%%%%%%%%%%%%%%%%%%%%%%%%%%%%%%%%%%%%%%%%%%%%%%%%%%%%
\section{Data Availability}
Certain data used in the preparation of this article were obtained from the Alzheimer’s
Disease Neuroimaging Initiative (ADNI) database. The ADNI was launched in 2003 as a
public-private partnership, led by Principal Investigator Michael W. Weiner, MD. The
primary goal of ADNI has been to test whether serial magnetic resonance imaging (MRI),
positron emission tomography (PET), other biological markers, and clinical and
neuropsychological assessment can be combined to measure the progression of mild
cognitive impairment (MCI) and early Alzheimer’s disease (AD).

Certain data used in the preparation of this article were obtained from the Critical Path
for Alzheimer's Disease (CPAD) database. In 2008, Critical Path Institute, in collaboration
with the Engelberg Center for Health Care Reform at the Brookings Institution, formed the
Coalition Against Major Diseases (CAMD), which was then renamed to CPAD in 2018. The
Coalition brings together patient groups, biopharmaceutical companies, and scientists from
academia, the U.S. Food and Drug Administration (FDA), the European Medicines Agency
(EMA), the National Institute of Neurological Disorders and Stroke (NINDS), and the
National Institute on Aging (NIA). CPAD currently includes over 200 scientists, drug
development and regulatory agency professionals, from member and non-member
organizations. The data available in the CPAD database has been volunteered by CPAD
member companies and non-member organizations.

Data used in the preparation of this article from the Quinn et al 2010 study were obtained from the University of
California, San Diego Alzheimer’s Disease Cooperative Study legacy database.

\subsection{Acknowledgements}
Data collection and sharing for this project was funded in part by the Alzheimer's Disease Neuroimaging Initiative (ADNI) (National Institutes of Health Grant U01 AG024904) and DOD ADNI (Department of Defense award number W81XWH-12-2-0012). ADNI is funded by the National Institute on Aging, the National Institute of Biomedical Imaging and Bioengineering, and through generous contributions from the following: AbbVie, Alzheimer’s Association; Alzheimer’s Drug Discovery Foundation; Araclon Biotech; BioClinica, Inc.; Biogen; Bristol-Myers Squibb Company; CereSpir, Inc.; Cogstate; Eisai Inc.; Elan Pharmaceuticals, Inc.; Eli Lilly and Company; EuroImmun; F. Hoffmann-La Roche Ltd and its affiliated company Genentech, Inc.; Fujirebio; GE Healthcare; IXICO Ltd.; Janssen Alzheimer Immunotherapy Research \& Development, LLC.; Johnson \& Johnson Pharmaceutical Research \& Development LLC.; Lumosity; Lundbeck; Merck \& Co., Inc.; Meso Scale Diagnostics, LLC.; NeuroRx Research; Neurotrack Technologies; Novartis Pharmaceuticals Corporation; Pfizer Inc.; Piramal Imaging; Servier; Takeda Pharmaceutical Company; and Transition Therapeutics. The Canadian Institutes of Health Research is providing funds to support ADNI clinical sites in Canada. Private sector contributions are facilitated by the Foundation for the National Institutes of Health (\href{url}{www.fnih.org}). The grantee organization is the Northern California Institute for Research and Education, and the study is coordinated by the Alzheimer’s Therapeutic Research Institute at the University of Southern California. ADNI data are disseminated by the Laboratory for Neuro Imaging at the University of Southern California.

\subsection{Financial Disclosure}
AMV, JLR, and DPM are equity-holding employees of, and AS a consultant for, Unlearn.ai, Inc., a company that creates software for clinical research and has patents pending for work described (US 17/808,954) and referenced (17/074,364) herein. Some companies from the biopharmaceutical industry have donated to ADNI (see above) and/or served as CPAD member organizations (\href{url}{https://c-path.org/programs/cpad/}).

\bibliographystyle{apalike}
\bibliography{refs}

\begin{thebibliography}{}

\bibitem[Bertolini et~al., 2020]{Bertolini}
Bertolini, D., Loukianov, A.~D., Smith, A.~M., Li-Bland, D., Pouliot, Y.,
  Walsh, J.~R., and Fisher, C.~K. (2020).
\newblock {Modeling Disease Progression in Mild Cognitive Impairment and
  Alzheimer's Disease with Digital Twins}.
\newblock {\em arXiv}.
\newblock https://arxiv.org/pdf/2012.13455.pdf.

\bibitem[Borm et~al., 2007]{Borm}
Borm, G.~F., Fransen, J., and Lemmens, W.~A. (2007).
\newblock {A simple sample size formula for analysis of covariance in
  randomized clinical trials}.
\newblock {\em Journal of Clinical Epidemiology}, 60(12):1234--1238.

\bibitem[DiMasi et~al., 2016]{DiMasi}
DiMasi, J.~A., Grabowski, H.~G., and Hansen, R.~W. (2016).
\newblock {Innovation in the pharmaceutical industry: New estimates of R\&D
  costs}.
\newblock {\em Journal of Health Economics}, 47:20--33.

\bibitem[{European Medicines Agency Committee for Medicinal Products for Human
  Use (CHMP)}, 2015]{emaCovar}
{European Medicines Agency Committee for Medicinal Products for Human Use
  (CHMP)} (2015).
\newblock Guideline on adjustment for baseline covariates in clinical trials.
\newblock
  https://www.ema.europa.eu/en/documents/scientific-guideline/guideline-adjustment-baseline-covariates-clinical-trials\_en.pdf.

\bibitem[{European Medicines Agency Committee for Medicinal Products for Human
  Use (CHMP)}, 2022]{EMA_2022}
{European Medicines Agency Committee for Medicinal Products for Human Use
  (CHMP)} (2022).
\newblock {Qualification} opinion for {Prognostic} {Covariate} {Adjustment}
  ({PROCOVA}™).
\newblock
  https://www.ema.europa.eu/en/documents/regulatory-procedural-guideline/qualification-opinion-prognostic-covariate-adjustment-procovatm\_en.pdf.

\bibitem[{Food and Drug Administration} et~al., 2021]{FDA_Covar}
{Food and Drug Administration}, {US Department of Health and Human Services},
  {Center for Drug Evaluation and Research (CDER)}, and {Center for Biologics
  Evaluation and Research (CBER)} (2021).
\newblock Adjusting for {Covariates} in {Randomized} {Clinical} {Trials} for
  {Drugs} and {Biological} {Products}: {Draft} {Guidance} for {Industry}.
\newblock
  https://www.fda.gov/regulatory-information/search-fda-guidance-documents/adjusting-covariates-randomized-clinical-trials-drugs-and-biological-products.

\bibitem[Ge et~al., 2011]{Ge}
Ge, M., Durham, L.~K., Meyer, R.~D., Xie, W., and Thomas, N. (2011).
\newblock {Covariate-Adjusted Difference in Proportions from Clinical Trials
  Using Logistic Regression and Weighted Risk Differences}.
\newblock {\em Drug information journal : DIJ / Drug Information Association},
  45(4):481--493.

\bibitem[Ghadessi et~al., 2020]{Ghadessi}
Ghadessi, M., Tang, R., Zhou, J., Liu, R., Wang, C., Toyoizumi, K., Mei, C.,
  Zhang, L., Deng, C.~Q., and Beckman, R.~A. (2020).
\newblock {A roadmap to using historical controls in clinical trials – by
  Drug Information Association Adaptive Design Scientific Working Group
  (DIA-ADSWG)}.
\newblock {\em Orphanet Journal of Rare Diseases}, 15(1):69.

\bibitem[Gneiting and Walz, 2021]{gneiting}
Gneiting, T. and Walz, E.-M. (2021).
\newblock {Receiver operating characteristic (ROC) movies, universal ROC (UROC)
  curves, and coefficient of predictive ability (CPA)}.
\newblock {\em Machine Learning}, pages 1--29.

\bibitem[Greenland and Robins, 1985]{Greenland}
Greenland, S. and Robins, J.~M. (1985).
\newblock {Estimation of a common effect parameter from sparse follow-up data.}
\newblock {\em Biometrics}, 41(1):55--68.

\bibitem[Hernández et~al., 2004]{Hernandez}
Hernández, A.~V., Steyerberg, E.~W., and Habbema, J.~F. (2004).
\newblock {Covariate adjustment in randomized controlled trials with
  dichotomous outcomes increases statistical power and reduces sample size
  requirements}.
\newblock {\em Journal of Clinical Epidemiology}, 57(5):454--460.

\bibitem[King and Nielsen, 2019]{king_nielsen_2019}
King, G. and Nielsen, R. (2019).
\newblock Why propensity scores should not be used for matching.
\newblock {\em Political Analysis}, 27(4):435–454.

\bibitem[Neuhaus, 1998]{Neuhaus}
Neuhaus, J.~M. (1998).
\newblock {Estimation Efficiency With Omitted Covariates in Generalized Linear
  Models}.
\newblock {\em Journal of the American Statistical Association},
  93(443):1124--1129.
\newblock In testing the null hypothesis, omitting covariate (when covariate \&
  treatment arm are independent) always looses efficiency; the stronger the
  association between omitted covariate and outcome is, the more the loss.

\bibitem[Noma and Nagashima, 2016]{Noma}
Noma, H. and Nagashima, K. (2016).
\newblock {A Note on the Mantel-Haenszel Estimators When the Common Effect
  Assumptions Are Violated}.
\newblock {\em Epidemiologic Methods}, 5(1):19--35.

\bibitem[Quinn et~al., 2010]{Quinn}
Quinn, J.~F., Raman, R., Thomas, R.~G., Yurko-Mauro, K., Nelson, E.~B., Dyck,
  C.~V., Galvin, J.~E., Emond, J., Jack, C.~R., Weiner, M., Shinto, L., and
  Aisen, P.~S. (2010).
\newblock {Docosahexaenoic Acid Supplementation and Cognitive Decline in
  Alzheimer Disease: A Randomized Trial}.
\newblock {\em JAMA}, 304(17):1903--1911.

\bibitem[Richardson et~al., 2017]{Richardson}
Richardson, T.~S., Robins, J.~M., and Wang, L. (2017).
\newblock {On Modeling and Estimation for the Relative Risk and Risk
  Difference}.
\newblock {\em Journal of the American Statistical Association},
  112(519):1121--1130.

\bibitem[Rockwood et~al., 2007]{Rockwood}
Rockwood, K., Fay, S., Gorman, M., Carver, D., and Graham, J.~E. (2007).
\newblock {The clinical meaningfulness of ADAS-Cog changes in Alzheimer's
  disease patients treated with donepezil in an open-label trial}.
\newblock {\em BMC Neurology}, 7(1):26.

\bibitem[Schuler et~al., 2021]{Schuler}
Schuler, A., Walsh, D., Hall, D., Walsh, J., Fisher, C., Disease, C. P. f.~A.,
  Initiative, A. D.~N., and Study, A. D.~C. (2021).
\newblock {Increasing the efficiency of randomized trial estimates via linear
  adjustment for a prognostic score}.
\newblock {\em The International Journal of Biostatistics}.

\bibitem[Siegfried et~al., 2022]{Siegfried}
Siegfried, S., Senn, S., and Hothorn, T. (2022).
\newblock {On the relevance of prognostic information for clinical trials: A
  theoretical quantification}.
\newblock {\em Biometrical Journal}.

\bibitem[Thorlund et~al., 2020]{thorlund}
Thorlund, K., Dron, L., Park, J. J.~H., and Mills, E.~J. (2020).
\newblock {Synthetic and External Controls in Clinical Trials – A Primer for
  Researchers}.
\newblock {\em Clinical Epidemiology}, 12:457--467.

\bibitem[Viechtbauer, 2010]{metafor}
Viechtbauer, W. (2010).
\newblock Conducting meta-analyses in r with the metafor package.
\newblock {\em Journal of Statistical Software}, 36(3):1--48.

\bibitem[Wong et~al., 2019]{Wong}
Wong, C.~H., Siah, K.~W., and Lo, A.~W. (2019).
\newblock {Estimation of clinical trial success rates and related parameters}.
\newblock {\em Biostatistics}, 20(2):273--286.

\end{thebibliography}

\renewcommand{\thesubsection}{\Alph{subsection}}
\pagebreak
\section{Appendix}

\subsection{Parameter notation}\label{appendix:notation-table}

\begin{table}[!h]
\centering
\resizebox{\columnwidth}{!}{
\begin{tabular}{ll}
 $n$ & Trial sample size. \\
 $Y$ & Binary outcome $y \in \{0,1\}$. \\
 $M$ & Prognostic score (probability of outcome) as produced by a prognostic model using historical data. \\
 $X$ & Stratifying covariate defined by a function of the prognostic scores $M$. \\
 $\boldsymbol{v} = \{0,v_1,...,v_{J-1}, 1\}$ & Set of cutoff values that define the strata, defined from the set $\boldsymbol{M}$. \\
 $A$ & Indicator for control ($a=0$) or treatment ($a=1$) arm assignment. \\
 $\boldsymbol{D}$ & Dataset of size $n$ comprising of $(X_i, A_i, Y_i)$, $i \in 1,...,n$. \\
 $J$ & Number of strata. \\
 $\pi_a$ & Randomization probability to arm $a$, ($\pi_0 = 1 - \pi_1$). \\
 ${\mu}_a$, $\hat{\mu}_a$, $\Tilde{\mu}_a$ & Marginal population, sampled, and modeled outcome probabilities for treatment $a$, respectively. \\
 $\mu_{aj}$, $\hat{\mu}_{aj}$, $\Tilde{\mu}_{aj}$ & Population, sampled, and modeled outcome probabilities on treatment $a$ in stratum $j$. \\
  $\phi_j$, $\varphi_j$ & Population and model strata propensities, or assignment probabilities. \\
 ${\psi}$, $\psi_j$ & Marginal risk ratio and risk ratio on stratum $j$, respectively. \\
 $\sigma^2$, $\hat{\sigma}^2$ & Asymptotic sampling variance and observed sample variance of $\log[\psi]$. \\
 %$\varsigma^2$ & Modeled asymptotic variance using modeled probabilities $u$. \\
 %$w_j$ & Mantel-Haenszel weight for stratum $j$. \\
 $r_{XY}$ & Spearman correlation between $X$ and $Y$. \\
 $*$ & Denotes historical data for any of the above parameters. \\
 $\gamma$ & Variance reduction. \\
 %$\xi$ & Sample size reduction. 
\end{tabular}
}
\caption{Parameter notation.}
\label{tab:notation}
\end{table}

\subsection{Derivation of the asymptotic variance estimator} \label{appendix:noma-derivation}

\cite{Noma} give us the asymptotic sampling variance estimator of the \textit{risk ratio}
\begin{equation*}
    \label{eq:noma-esteq}
    \Var_\infty[{\psi}] = \frac{\sum_j (\phi_j\pi_1\pi_0)^2[\text{Var}(g(\psi_j))]}{\left(\sum_j \phi_j\pi_1\pi_0\mu_{0j}\right)^2}
\end{equation*}
where $g(\psi) = \mu_{1j} - \psi_j\mu_{0j}$. To solve for $\text{Var}(g(\psi_j))$, we use the Delta Method. The outline below is given for some stratum $j$ and follows for the other $J-1$ strata.

Consider the parameter vector $[\pi_0\mu_0, \pi_1\mu_1, \pi_0, \pi_1]$, such that formally $g(U) = \frac{\pi_1\mu_1}{\pi_1} - \psi\frac{\pi_0\mu_0}{\pi_0}$. Then
\begin{equation*}
    \text{Var}_\infty(g(U)) = \nabla g(\tau)^\top \text{V}_\infty(U) \nabla g(\tau)
\end{equation*}
where $\tau = \E[U]$, $\nabla g(\tau)$ is the gradient of $g$, and $\text{V}_\infty(U)$ is the variance-covariance matrix for $U$. With some algebra, the asymptotic distribution of $U$ is given by
\begin{equation*}
    \E[U] = \phi \begin{bmatrix} \pi_0\mu_0 \\ \pi_1\mu_1 \\ \pi_0 \\ \pi_1 \end{bmatrix}
\end{equation*}
\begin{equation*}
    \text{V}_\infty(U) = -\phi 
    \begin{bmatrix} \pi_0\mu_0(\phi\pi_0\mu_0 - 1) & \phi\pi_0\pi_1\mu_0\mu_1 & \pi_0\mu_0(\phi\pi_0 - 1) & \phi\pi_0\pi_1\mu_0 \\
    & \pi_1\mu_1(\phi\pi_1\mu_1 - 1) & \phi\pi_0\pi_1\mu_1 & \pi_1\mu_1(\phi\pi_1 - 1) \\ 
    & & \pi_0(\phi\pi_0 - 1) & \phi\pi_0\pi_1 \\
    & & & \pi_1(\phi\pi_1 - 1) 
    \end{bmatrix}
\end{equation*}
Additionally, the gradient is given by
\begin{equation*}
    \nabla g(\tau) = -\phi \begin{bmatrix} -\psi/\pi_0 \\ 1/\pi_1 \\ \psi\mu_0/\pi_0 \\ -\mu_1/\pi_1 \end{bmatrix}
\end{equation*}
Putting it together, we have 
\begin{equation*}
    \text{Var}_\infty(g(U)) = \psi^2 \left(\frac{\mu_{0}(1-\mu_{0})}{\pi_0}\right) + \frac{\mu_{1}(1-\mu_{1})}{\pi_1}
\end{equation*}
Finally, we get the asymptotic variance estimator for $\bar{\psi}$:
\begin{equation*}
    \label{eq:noma-delta-method}
    \Var_\infty[{\psi}] = \frac{\sum_j \phi_j (\pi_0 \pi_1)^2[\psi_j^2 \frac{\mu_{0j}(1-\mu_{0j})}{\pi_0} + \frac{\mu_{1j}(1-\mu_{1j})}{\pi_1}]}{\left(\sum_j \phi_j \pi_0 \pi_1 \mu_{0j}\right)^2}
\end{equation*}
With a bit of basic algebra, this simplifies to
\begin{equation*}
    \Var_\infty[{\psi}] = \frac{\sum_j \phi_j\psi_j[\pi_0{\mu}_{0j} + \pi_1\mu_{1j} - \mu_{0j}\mu_{1j}]}{\pi_1\pi_0\left(\sum_j \phi_j\mu_{0j}\right)^2}
\end{equation*}
with $\bar{\mu}_j = \pi_0 \mu_{0j} + \pi_1 \mu_{1j}$. Using the Delta method, we get, for the log risk ratio, $\sqrt{n}\Var[\log({\psi})] \xrightarrow[]{D} \mathcal{N}\left(\log({\psi}), \frac{1}{\psi^2}\Var[{\psi}]\right)$:
\begin{equation*}
\label{eq:delta-var-logrr}
    \text{Var}_\infty[\sqrt{n}\log({\psi})] = \sigma_\infty^2 = \frac{\sum_j \phi_j\psi_j[\pi_0{\mu}_{0j} + \pi_1\mu_{1j} - \mu_{0j}\mu_{1j}]}{\psi^2\pi_1\pi_0\left(\sum_j \phi_j\mu_{0j}\right)^2}
\end{equation*}

%Greenland and Robins give an estimator for the sampling variance under an assumption of homoegeneous treatment effect that is consistent for Eq. \ref{eq:delta-var-logrr} in large samples \cite{Greenland}.

\subsection{Using the plug-in estimator to prospectively estimate asymptotic sample variance.} \label{appendix:proc-plugin}

\begin{algorithm}[H]
    \caption{Using the plug-in estimator for prospectively estimate asymptotic sample variance.}
  \begin{algorithmic}[1]
    \State Use Procedure \ref{proc:define-strata} to obtain estimates of $\boldsymbol{\hat{\mu}} = \{\hat{\mu}_{01},...,\hat{\mu}_{0J} \}$ and $\boldsymbol{\hat{\phi}^{}} = \{\hat{\phi}_1,...,\hat{\phi}_J \}$ from historical data $\boldsymbol{D^*}$.
    \State Specify trial parameters risk ratio ${\psi}$ and randomization probability $\pi_1$.
    \State Plug all values into Eq. \ref{eq:hist-plugin-var} to estimate the asymptotic sampling variance of the treatment effect.
  \end{algorithmic}
\label{proc:prospective-var-plugin}
\end{algorithm}

\subsection{Variance reduction} \label{appendix:var_ratio}

For an unadjusted analysis, the asymptotic variance is
\begin{equation*}
     \sigma^2_{unadjusted} = \frac{{\psi}\left[\pi_0{\mu}_{0} + \pi_1\mu_{1} - {\mu}_0{\mu}_1\right]}{\pi_0 \pi_1 {\mu}_0^2}
\end{equation*}
Plugging into $\gamma = 1 - \frac{\sigma^2_{{PROCOVA-CMH}}}{\sigma^2_{unadjusted}}$ gives
\begin{align*}
    \gamma &= 1 - \left( \frac{\sum_j \phi_j \psi_j[\pi_0{\mu}_{0j} + \pi_1\mu_{1j} - \mu_{0j}\mu_{1j}]}{\pi_0 \pi_1\left(\sum_j \phi_j \mu_{0j}\right)^2} \right) \left( \frac{\pi_0 \pi_1 {\mu}_0^2}{{\psi}\left[\pi_0{\mu}_{0} + \pi_1\mu_{1} - {\mu}_0{\mu}_1\right]} \right) \\
    %&= 1 - \frac{(\pi_0 \pi_1)^2 {\mu}_0^2 \sum_j \phi_j \psi_j [\pi_0{\mu}_{0j} + \pi_1\mu_{1j} - \mu_{0j}\mu_{1j}]}{(\pi_0\pi_1)^2{\psi}\left[\pi_0{\mu}_{0} + \pi_1\mu_{1} - {\mu}_0{\mu}_1\right] \left(\sum_j \phi_j\mu_{0j}\right)^2} \\
    &= 1 - \frac{{\mu}_0^2 \sum_j \phi_j\psi_j[\pi_0{\mu}_{0j} + \pi_1\mu_{1j} - \mu_{0j}\mu_{1j}]}{{\psi}\left[\pi_0{\mu}_{0} + \pi_1\mu_{1} - {\mu}_0{\mu}_1\right] \left(\sum_j {\phi}_j{\mu}_{0j}\right)^2}
\end{align*}
The values used for the parameters (e.g. $\hat{\mu}$ or $\Tilde{\mu}$) depend on which historically-informed prospective variance estimator (plug-in (Eq. \ref{eq:hist-plugin-var}) or modeled (Eq. \ref{eq:var-modeled})) is used.

%Note that because we are operating under a balanced design, the randomization probabilities are equal across strata, and thus cancel.

%\subsection{Sample size reduction} \label{appendix:N-reduction}

\end{document}